\documentclass[times,doublespace]{fldauth}
\usepackage{lineno,hyperref}
\usepackage{bm, amsmath, esint, mathtools,xcolor}
\usepackage{graphicx}
\usepackage{caption,subcaption}
\begin{document}
\runningheads{Verma et al.}{Surface-forces on deforming geometries}
\title{Computing the Force Distribution on the Surface of Complex, Deforming Geometries using Vortex Methods and Brinkman Penalization}

\author{Siddhartha Verma\affil{1}, Gabriele Abbati\affil{1}, Guido Novati\affil{1}\ and Petros Koumoutsakos\affil{1}\corrauth}
\address{\affilnum{1}Computational Science and Engineering Laboratory, Clausiusstrasse 33, ETH Z{\"u}rich, CH-8092, Switzerland}
\corraddr{Computational Science and Engineering Laboratory, Clausiusstrasse 33, ETH Z{\"u}rich, CH-8092, Switzerland (Tel.: +41 44 632 71 59; fax: +41 44 632 17 03). E-mail: petros@ethz.ch}
\cgs{European Research Council Advanced Investigator Award}
\cgsn{Swiss National Science Foundation}{CRSII3\_147675}

\begin{abstract}
The distribution of forces on the surface of complex, deforming geometries is an invaluable output of flow simulations. One particular example of such geometries involves self-propelled swimmers. Surface forces can provide significant information about the flow field sensed by the swimmers, and are difficult to obtain experimentally. At the same time, simulations of flow around complex, deforming shapes can be computationally prohibitive when body-fitted grids are used. Alternatively, such simulations may employ penalization techniques. Penalization methods rely on simple Cartesian grids to discretize the governing equations, which are enhanced by a penalty term to account for the boundary conditions. They have been shown to provide a robust estimation of mean quantities, such as drag and propulsion velocity, but the computation of surface force distribution remains a challenge. We present a method for determining flow-induced forces on the surface of both rigid and deforming bodies, in simulations using re-meshed vortex methods and Brinkman penalization. The pressure field is recovered from the velocity by solving a Poisson's equation using the Green's function approach, augmented with a fast multipole expansion and a tree-code algorithm. The viscous forces are determined by evaluating the strain-rate tensor on the surface of deforming bodies, and on a `lifted' surface in simulations involving rigid objects. We present results for benchmark flows demonstrating that we can obtain an accurate distribution of flow-induced surface-forces. The capabilities of our method are demonstrated using simulations of self-propelled swimmers, where we obtain the pressure and shear distribution on their deforming surfaces.  
\end{abstract}

\maketitle

%\linenumbers

%%%%%%%%%%%%%%%%%%%%
\section{Introduction}
The distribution of surface-forces, and its dependence on the shape and motion of biological organisms and engineering devices, is among the most valuable information that can be acquired through numerical simulations. Such information may not be easy to obtain experimentally, particularly for biological organisms, such as swimmers that involve complex deforming bodies. The measurement of surface-forces was the subject of some of the pioneering observations concerned with swimming organisms by Gray~\cite{Gray1957}, and experiments by DuBois \& Ogilvy~\cite{Dubois1978}. Theoretical studies of fish-swimming by Lighthill~\cite{Lighthill1970} and Wu~\cite{Wu1971} employed potential flow theory to provide estimates for the pressure field on the body-surface. More recent work~\cite{Dabiri2005,Dabiri2014} has provided new impetus for the measurement of such quantities, which are deemed to be essential for answering some of the most critical questions concerning our understanding of fish-swimming~\cite{Lauder2009}. 

Simulations can complement experiments by providing complete information regarding the distribution of surface-forces on swimming organisms. Direct Numerical Simulations of swimming organisms were first performed using Arbitrary Lagrangian-Eulerian (ALE) methods~\cite{Liu1996,Liu1997,Carling1998}. ALE methods~\cite{Hirt1974, Hughes1981, Liu1999, Kern2006} utilize body-fitted grids, where the surface of the solid is delineated explicitly by the mesh. This makes enforcing boundary-conditions relatively straightforward. However, body-fitted grids may encounter difficulties with bodies experiencing large deformations, and the grid must be reconstructed frequently to account for moving and deforming objects. Frequent mesh regeneration is undesirable, as it entails significant computational overhead. This overhead can be especially prohibitive when performing shape and motion optimisation of fish-swimming using evolutionary algorithms~\cite{Kern2006}. Alternative approaches, referred to as Immersed Boundary (IB) methods~\cite{Peskin1972,Peskin2002,Mittal2005,Gilmanov2005,Hieber2008,Borazjani2009}, reduce computational cost by using simple, non-body-fitted meshes. The no-slip condition is imposed on curved boundaries using appropriate interpolation of the velocity field across the fluid-solid interface. In a different approach, Cartesian grids have been combined with penalisation techniques~\cite{Angot1999} for simulations of fish-swimming~\cite{Gazzola2011,Bergmann2011}. Penalisation techniques allow for easy computation of the rotational and translational motion of self-propelled swimmers, via integrals of the flow field over the body volume. However, penalisation techniques present challenges when detailed distributions of the pressure-induced and viscous forces are required on the body surface. Such force fields may be noisy owing to the irregular intersection of the curved body-geometry with the Cartesian grid.
 
Early developments of Immersed Boundary methods were coupled with vortex methods~\cite{McCracken1980} for enforcing the no-slip boundary condition. Vortex methods primarily discretize the vorticity present in the flow field, and can prove to be advantageous in flows involving a limited support of the vorticity field, such as fish wakes. These early simulations were abandoned due to inaccuracies of vortex methods at that time. Such limitations were overcome in later work that employed re-meshing to enforce the regularisation of particle locations~\cite{Koumoutsakos1997}, or utilized a vorticity flux algorithm for the enforcement of the no-slip condition~\cite{Koumoutsakos1994}, and made use of variable kernel sizes and domain-decomposition techniques to account for multiple bodies~\cite{Cottet2000}. However, their applicability to 3D and multiple deforming bodies was limited due to the requirement of body fitted grids, or due to limitations imposed on the Cartesian meshes by remeshing. 

Recent work utilizing the combination of vortex methods with Brinkman penalization has demonstrated that the approach is extremely effective in simulating fluid-structure interaction of complex, temporally deforming geometries~\cite{Coquerelle2008,Gazzola2011, Rossinelli2015}. A number of studies have relied on this technique for coupling simulations of self-propelled swimmers with machine-learning algorithms~\cite{Gazzola2014}, and for design optimization of swimmer morphology and kinematics~\cite{vanRees2013,Gazzola2012,Gazzola2012PoF,vanRees2015}. These studies demonstrate the capability of the penalization algorithm to handle multiple, complex and deforming geometries with relative ease. Nonetheless, these studies have not overcome the difficulty of determining pointwise forces exerted on a solid surface. Surface-forces are the primary means through which fluids exert an influence on solid objects, and knowledge of these forces is essential for analyzing fluid-solid interactions. The absence of an explicit calculation for the pressure term in vortex methods, and the smooth interface used for the penalization algorithm, pose considerable challenges when determining surface-forces. These issues are overcome in the current work. A complete methodology is presented to accurately compute surface-forces in simulations of two-dimensional flows past complex deforming geometries, using Cartesian grids.

The paper is organised as follows. The equations for determining surface-forces, and the numerical techniques used, are described in Section~\ref{sec:numMeth}. Validations with simulations of impulsively started two-dimensional cylinders, simulations of flow over a rigid streamlined profile, and simulations of a self-propelled swimmer, are discussed in Section~\ref{sec:results}. Section~\ref{sec:summary} summarizes the results and the numerical techniques outlined in the paper.

%%%%%%%%%%%%%%%%%%%%%%%%%%%%%%%

\section{Numerical Methods}
\label{sec:numMeth}
This work is concerned  with two-dimensional, viscous, incompressible flows past complex, rigid and deforming geometries. In this section, we provide a brief description of the equations governing the interaction between fluid-flow and solid objects, and the numerical procedure used to compute the pressure-induced and viscous forces on a solid.

\subsection{Vortex methods and Brinkman penalization}
The spatial and temporal evolution of the velocity field in our simulations is based on the incompressible Navier-Stokes equations:
\begin{eqnarray}
\nabla\cdot\bm{u} &=& 0\\
\dfrac{\partial \bm{u}}{\partial t} + (\bm{u}\cdot\nabla)\bm{u} &=& \dfrac{-\nabla P}{\rho} + \nu\nabla^2\bm{u}
\label{eq:NS}
\end{eqnarray}
The interaction between fluid-flow and solid objects is achieved by introducing a penalty term in the momentum equation (Brinkman penalization~\cite{Angot1999}), which enforces the no-slip boundary condition at the fluid-solid interface:
\begin{equation}
\dfrac{\partial \bm{u}}{\partial t} + (\bm{u}\cdot\nabla)\bm{u} = \dfrac{-\nabla P}{\rho} + \nu\nabla^2\bm{u} + \lambda\chi\left(\bm{u}_s-\bm{u}\right)
\label{eq:penalNS}
\end{equation}
Here, $\lambda$ is the penalization parameter, and $\chi$ is the characteristic function which represents the discretized solid on a Cartesian grid. Grid points with $\chi=0$ are occupied entirely by the fluid, and those with $\chi=1$ are occupied entirely by the solid. To minimize numerical oscillations, $\chi$ transitions smoothly from $0$ to $1$ within 2 grid-points at the fluid-solid interface, using a discrete representation of the Heaviside function~\cite{Towers2009}. The symbol $\bm{u}_s$ in Eq.~\ref{eq:penalNS} denotes the pointwise velocity of the discretized solid, and accounts for translation, rotation, and deformation of the body.

To solve the motion resulting from fluid-solid interaction, we use an open-source solver~\cite{Rossinelli2015} based on re-meshed vortex methods~\cite{Koumoutsakos1995}. These methods use the vorticity form of the momentum equation, which may be obtained by taking the curl of Eq.~\ref{eq:penalNS}:
\begin{equation}
\dfrac{\partial \bm{\omega}}{\partial t} + (\bm{u}\cdot\nabla)\bm{\omega} = \nu\nabla^2\bm{\omega} + \lambda\nabla\times\left(\chi\left(\bm{u}_s-\bm{u}\right)\right)
\label{eq:penalNSvort}
\end{equation}
In deriving Eq.~\ref{eq:penalNSvort}, we have used the fact that $\nabla\cdot\bm{u} = 0$ (incompressibility), and $\nabla\rho=0$, since we restrict our investigation to neutrally buoyant solids (i.e., $\rho_{solid} = \rho_{liquid}$). Furthermore, the vortex-stretching term ($\bm{\omega}\cdot\nabla\bm{u}$) is absent from the momentum equation for two-dimensional cases. A detailed description of the time-splitting steps involved in solving Eq.~\ref{eq:penalNSvort} may be found in~\cite{Gazzola2011,Rossinelli2015}.

%%%%%%%%%%%%%%%%%%%%%%%%%%%%%%%%%%
\subsection{Recovering pressure from the velocity field}
\label{sec:pressurePoisson}
Advancing the vorticity field in time using Eq.~\ref{eq:penalNSvort} requires the computation of velocity at every time step ($\nabla^2\bm{u}=-\nabla\times\bm{\omega}$). The pressure field can be recovered from the velocity by taking the divergence of the penalized momentum equation (Eq.~\ref{eq:penalNS}):
\begin{equation}
\begin{split}
\dfrac{\partial \left(\nabla\cdot\bm{u}\right)}{\partial t} + \left(\nabla\bm{u}^T:\nabla\bm{u}\right) + \left(\bm{u}\cdot\nabla\right)\left(\nabla\cdot\bm{u}\right) = \dfrac{-\nabla^2 P}{\rho} + \dfrac{1}{\rho^2}\nabla\rho\cdot\nabla P + \nu\nabla^2\left(\nabla\cdot\bm{u}\right) \\
 + \lambda\nabla\cdot\left(\chi\left(\bm{u}_s-\bm{u}\right)\right)
\label{eq:PoissonDerive}
\end{split}
\end{equation}
The colon operator in $\nabla\bm{u}^T:\nabla\bm{u}$ denotes the tensor dot product, which can be written in index notation as $u_{j,i}u_{i,j}$. Using the incompressibility condition ($\nabla\cdot\bm{u}=0$) and the fact that $\nabla\rho = 0$ (neutrally buoyant solid), Eq.~\ref{eq:PoissonDerive} simplifies to a Poisson's equation for the pressure term:
\begin{equation}
 \nabla^2 P = -\rho\left(\nabla\bm{u}^T:\nabla\bm{u}\right) + \rho\lambda\nabla\cdot\left(\chi\left(\bm{u}_s-\bm{u}\right)\right)
\label{eq:PoissonFinal}
\end{equation}
This Poisson's equation is solved using a fast tree-code algorithm based on multipole expansions~\cite{Greengard1987}, as described in Appendix~\ref{sec:PoissonNumerics}. The penalty term in this equation ($\rho\lambda\nabla\cdot\chi\left(\bm{u_s}-\bm{u}\right)$) accounts for the pressure-source contribution induced by fluid-solid interactions.

%%%%%%%%%%%%%%%%%%%%%%%%%
\subsection{Determining the surface-forces}
When using Brinkman penalization, the absence of a sharply defined fluid-solid interface poses the biggest obstacle in determining the forces exerted on a solid surface. To overcome this difficulty, we consider the pressure-induced (isotropic) and viscosity-induced (deviatoric) parts of the forces separately.

%%%%%%%%%%%%%%%%%%%%%%%%%
\subsubsection{Pressure-induced forces}

Pressure-induced forces act on a surface immersed in a fluid, even when both the fluid and the solid are at rest. Computing these forces ($\bm{F}_P$) requires knowledge of the pressure, the surface normal ($\bm{n}$), and the infinitesimal surface area ($dS$):
\begin{equation}
d\bm{F}_P = -P \bm{n} \ dS
\label{eq:Fpress}
\end{equation}
The surface coordinates ($\bm{x}_{surf}$) and normals defining the solid boundary are known precisely at each time step in simulations. These exact surface-coordinates ($\bm{x}_{surf}$), which may not necessarily coincide with the grid nodes, are used as target locations $\bm{z}$ for computing $P\left(\bm{x}_{surf}\right)$ using Eq.~\ref{eq:expandedFinal}. Evaluating the pressure at the body-surface coordinates instead of the entire computational domain reduces the cost of the Poisson solver significantly.

Once the pointwise pressure-induced surface-forces are known from Eq.~\ref{eq:Fpress}, their resultant on the entire body can be determined using a closed surface integral:
\begin{eqnarray}
\bm{F}_{P} = \oiint d\bm{F}_{P}= \oiint  -P\bm{n} \ dS
\label{eq:FpressNet}
\end{eqnarray}
$\bm{F}_{P}$ can be projected along the velocity vector to obtain the contribution of pressure-induced forces to thrust and drag. For stationary solid objects, the pressure-induced drag and the corresponding drag coefficient are defined as follows:
\begin{equation}
\left(Drag\right)_P = \dfrac{\bm{F}_{P}\cdot\bm{U}_\infty}{\|\bm{U}_\infty\|} \quad \text{and} \quad Cd_P = \dfrac{2 \left(Drag\right)_P}{\rho\|\bm{U}_\infty\|^2 A}
\label{eq:CdP}
\end{equation}
where $\bm{U}_\infty$ is the free-stream velocity, and $A$ represents the reference area ($A=\left(2\cdot\text{radius}\right)$ for two-dimensional cylinders).
%%%%%%%%%%%%%%%%%%%
\subsubsection{Viscous forces}
\label{sec:viscForces}
To compute viscous forces ($\bm{F}_\nu$) resulting from relative motion between the fluid and the solid, we use the pointwise strain-rate tensor $\bm{D}\left(\bm{u}\left(\bm{x}\right)\right)$:
\begin{equation}
d\bm{F}_\nu = 2\mu \bm{D}\cdot \bm{n} \ dS \quad \text{where} \quad \bm{D} = \dfrac{1}{2}\left(\nabla\bm{u} + \nabla\bm{u}^T\right)
\label{eq:Fvisc}
\end{equation}
Here, $\mu$ is the dynamic viscosity of the fluid. For simulations involving deforming objects (i.e., self-propelled swimmers), $\nabla\bm{u}$ in Eq.~\ref{eq:Fvisc} is evaluated at the surface of the solid. However, in the case of rigid objects, smoothing of the solid boundary on the computational grid leads to inaccurate estimation of velocity gradients. To overcome this difficulty, $\nabla\bm{u}$ is computed on a `lifted' body surface via nearest-neigbour interpolation. The lifted-surface is formed by moving the original solid surface outward along the surface-normal. Our tests indicate that a distance of $2h$ from the exact edge of rigid objects yields the best accuracy for determining $\nabla\bm{u}$. With the pointwise viscous forces known from Eq.~\ref{eq:Fvisc}, the resultant quantities $\bm{F}_{\nu}$, $\left(Drag\right)_\nu$, and $Cd_\nu$ may be determined as follows:
\begin{eqnarray}
\bm{F}_{\nu} &=& \oiint 2\mu \bm{D}\cdot \bm{n} \ dS
\label{eq:FviscNet}\\
\left(Drag\right)_\nu = \dfrac{\bm{F}_{\nu}\cdot\bm{U}_\infty}{\|\bm{U}_\infty\|} \quad &\text{and}& \quad Cd_\nu = \dfrac{2 \left(Drag\right)_\nu}{\rho\|\bm{U}_\infty\|^2 A}
\label{eq:CdNu}
\end{eqnarray}

%%%%%%%%%%%%%%%%
\section{Results}
\label{sec:results}
In this section, we present results obtained using the numerical procedure described in Section~\ref{sec:numMeth}. All the simulations discussed were run using an open-source 2D incompressible Navier Stokes solver~\cite{Rossinelli2015}. The penalization-parameter in Eq.~\ref{eq:penalNSvort} was set to $\lambda = 10^6$, and the multipole expansion in Eq.~\ref{eq:pressTruncated} was truncated to $p=12$.

Surface-force computations for rigid objects are validated first using data available in the literature for impulsively started cylinders. This is followed by simulations of rigid fish-like profiles, where the drag force computed using surface-forces is compared to the drag obtained using the penalization algorithm. Validation for deforming geometries (self-propelled swimmers) is done by comparing the unsteady acceleration computed using surface-forces, to the acceleration determined using the penalization algorithm. We emphasize that the penalization algorithm yields just the resultant-force acting on a body, whereas the surface-force computations provide a detailed picture of pointwise forces acting on the object's surface.
 
\subsection{Impulsively started cylinders}
\label{sec:resultsCyl}
\begin{figure}
        \centering
        \begin{subfigure}[b]{0.49\textwidth}
                \centering
                \includegraphics{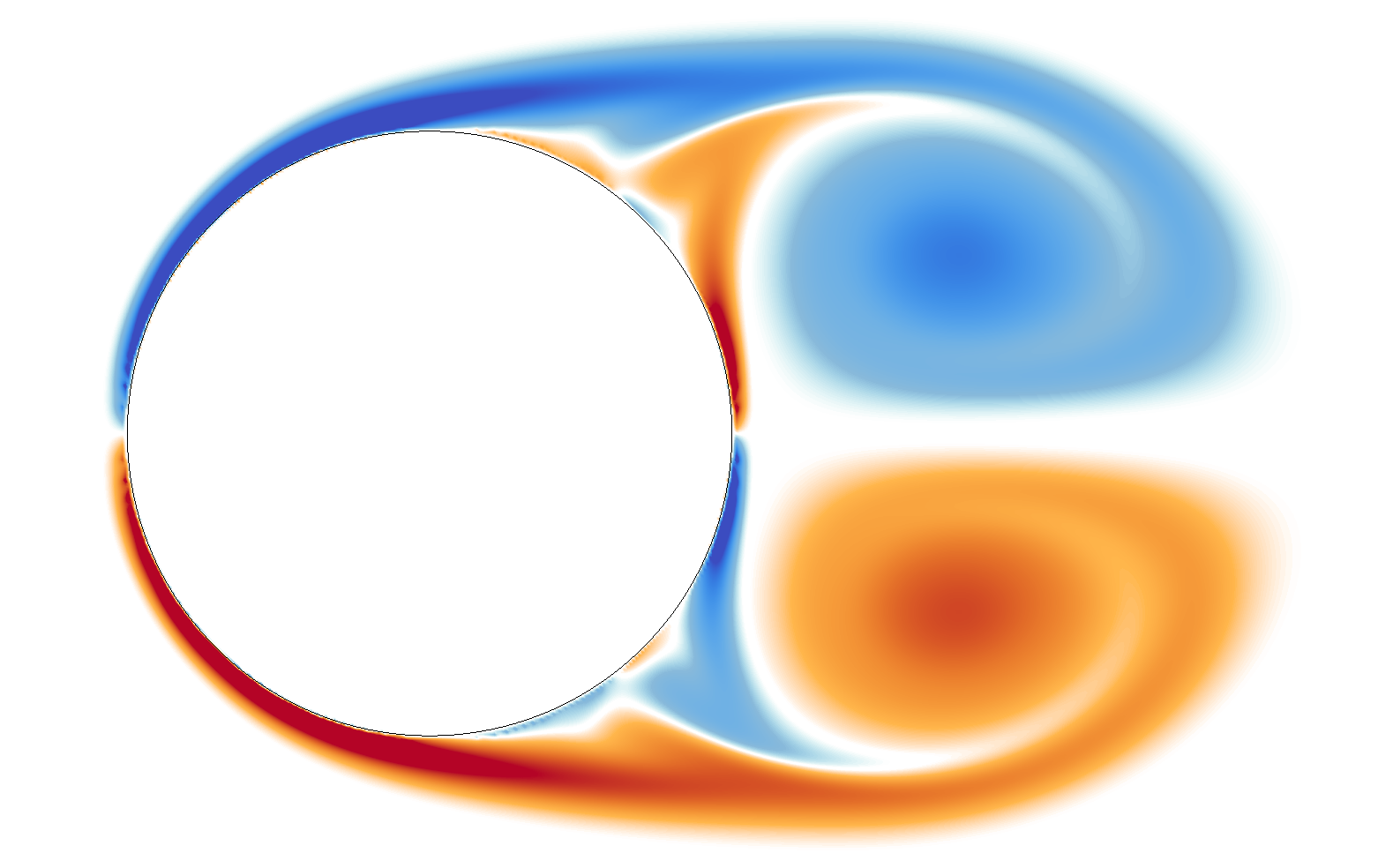}
                \subcaption{}
                \label{fig:cylVort}
        \end{subfigure}
        \begin{subfigure}[b]{0.49\textwidth}
                \centering
                \includegraphics{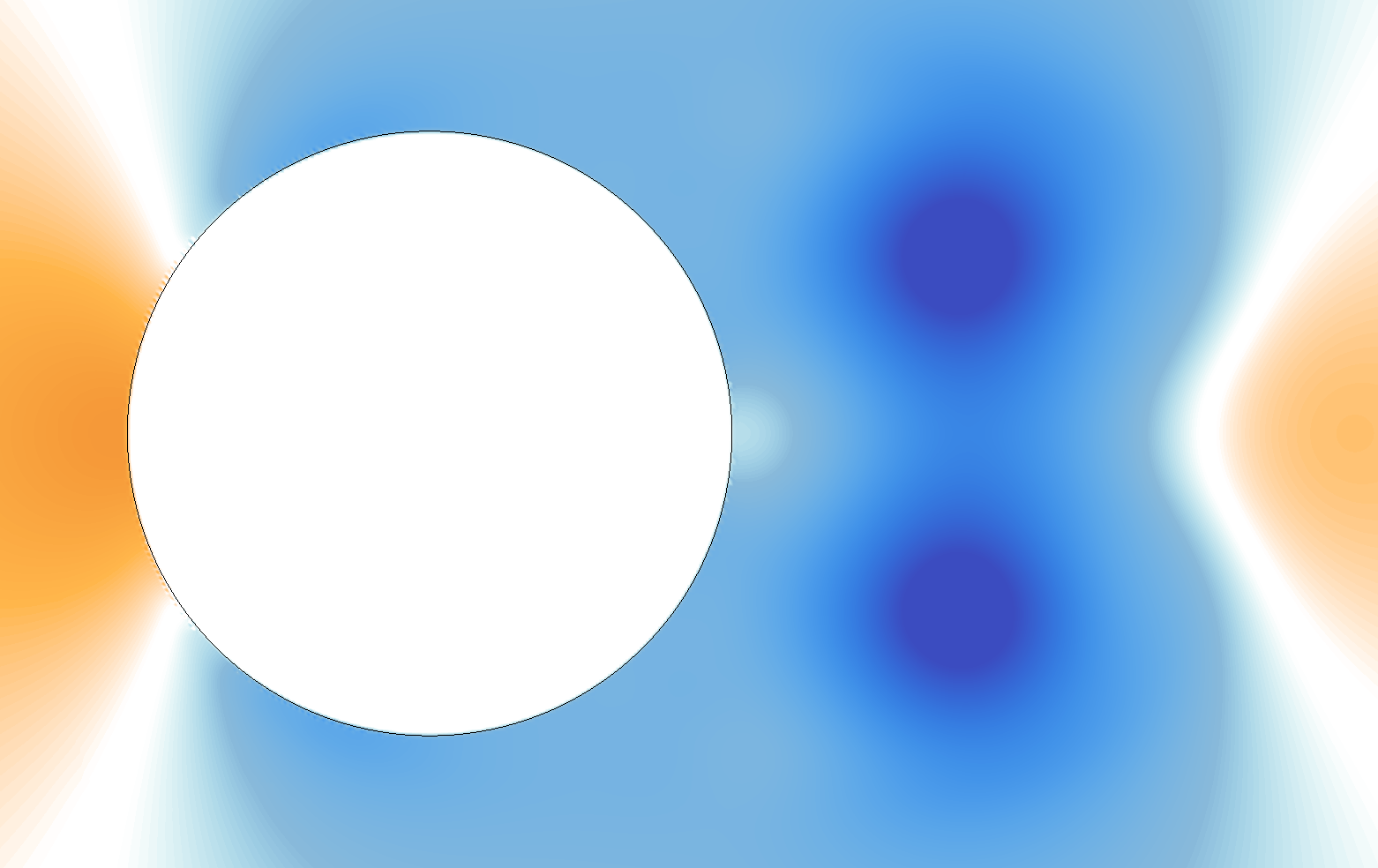}
                \subcaption{}
                \label{fig:cylPress}
        \end{subfigure}
  \caption{(\subref{fig:cylVort}) The vorticity field and (\subref{fig:cylPress}) the pressure field generated by flow over a stationary cylinder at $Re=1000$. Red regions correspond to positive vorticity and pressure, and blue regions correspond to negative values.}
\label{fig:cylVortPress}
\end{figure}

\begin{figure}
\centering
\includegraphics{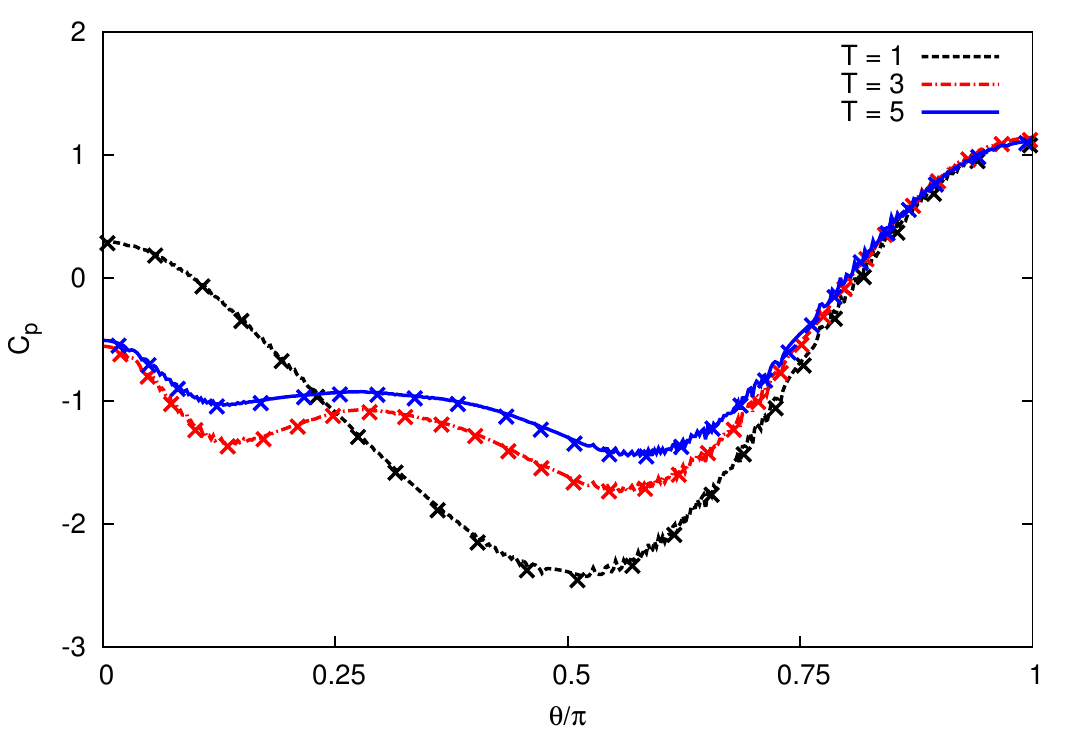}
\caption{Pressure coefficient ($C_P$) distribution on the cylinder's surface for $Re=550$, at three different non-dimensionalized times ($T=t\|\bm{U}_\infty\|/2R$). Lines: current work, symbols: reference data from~\cite{Li2004}.}
\label{fig:CpVsTheta}
\end{figure}

\begin{figure}
        \centering
        \begin{subfigure}[b]{\textwidth}
                \centering
                \begin{minipage}{.49\textwidth}
                \includegraphics{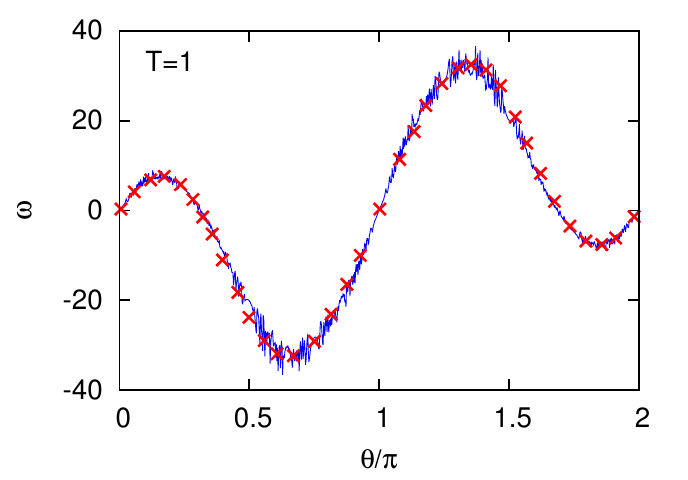}
                \end{minipage}
                \begin{minipage}{.49\textwidth}
                \includegraphics{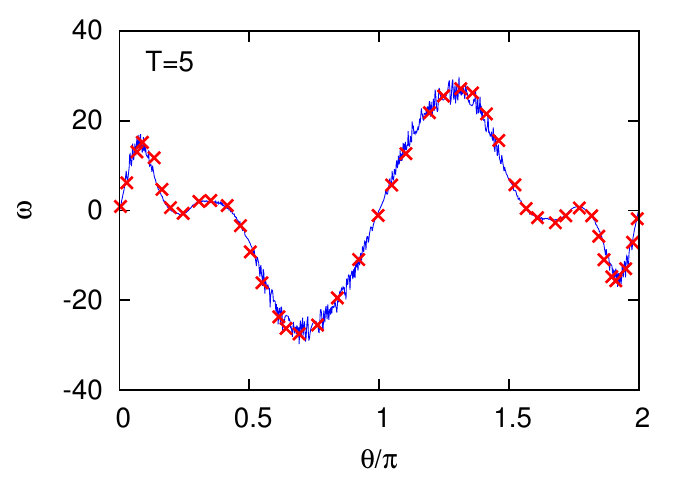}
                \end{minipage}
                \subcaption{}
                \label{fig:vortRe550}
        \end{subfigure}
        \begin{subfigure}[b]{\textwidth}
                \centering
                \begin{minipage}{.49\textwidth}
                \includegraphics{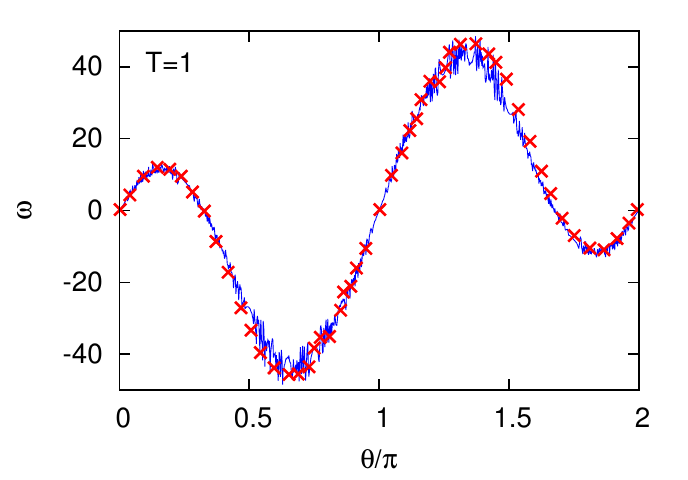}
                \end{minipage}
                \begin{minipage}{.49\textwidth}
                \includegraphics{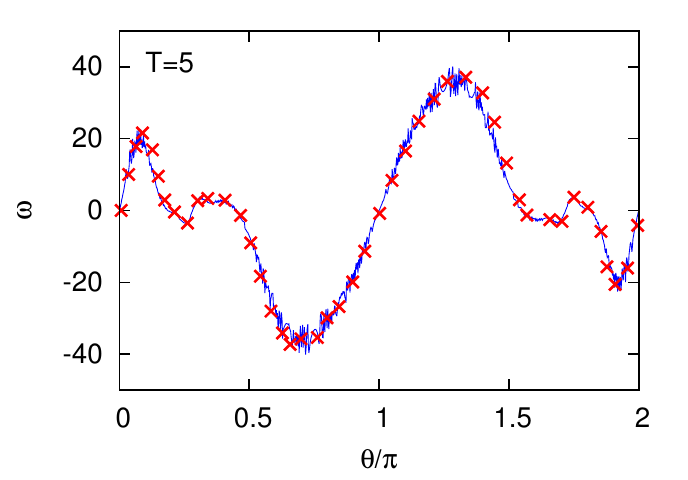}
                \end{minipage}
                \subcaption{}
                \label{fig:vortRe1000}
        \end{subfigure}
  \caption{Spatial distribution of non-dimensionalized vorticity computed on the lifted-surface of the cylinders at two time instances ($T=1$ and $T=5$), for (\subref{fig:vortRe550}) $Re=550$ and (\subref{fig:vortRe1000}) $Re=1000$. Lines represent current data, symbols correspond to reference data from~\cite{Koumoutsakos1995}.}
  \label{fig:vortCyl}
\end{figure}

\begin{figure}
        \centering
        \begin{subfigure}[b]{\textwidth}
                \centering
                \includegraphics{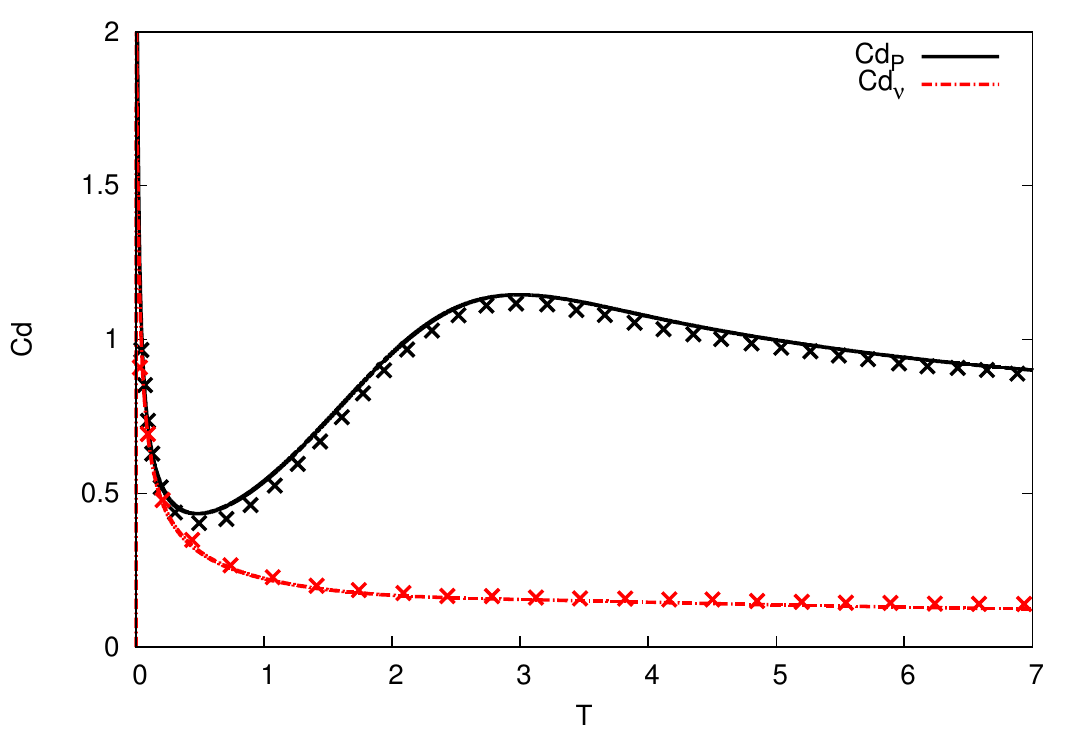}
                \subcaption{}
                \label{fig:CdRe550}
        \end{subfigure}
        \begin{subfigure}[b]{\textwidth}
                \centering
                \includegraphics{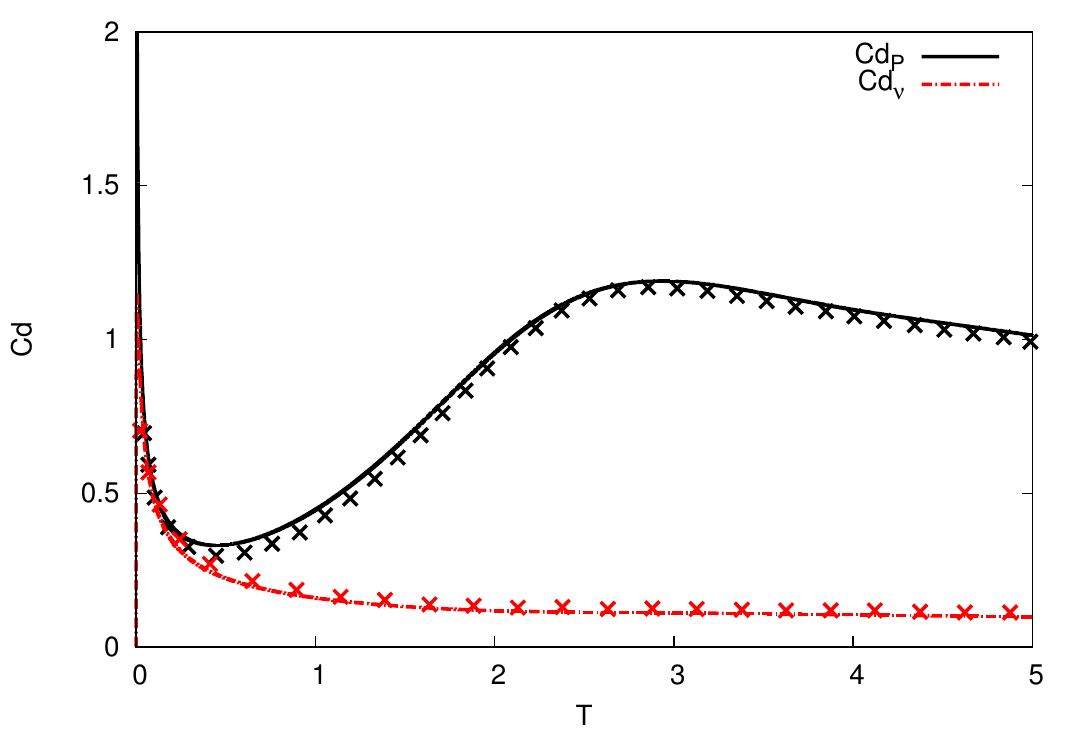}
                \subcaption{}
                \label{fig:CdRe1000}
        \end{subfigure}  \caption{Temporal evolution of pressure-drag coefficient ($Cd_P$) and viscous-drag coefficient ($Cd_\nu$) for impulsively started cylinders at (\subref{fig:CdRe550}) $Re=550$ and (\subref{fig:CdRe1000}) $Re=1000$. Lines represent data from current work, symbols correspond to reference data from~\cite{Koumoutsakos1995}.}
\label{fig:CdCylinder}
\end{figure}

We first examine impulsively started flow over rigid cylinders, for diameter-based Reynolds numbers of $Re=550$ and $1000$ ($Re = 2\|\bm{U}_\infty\| R/\nu$). Both these simulations were run in a unit square domain using a uniform grid of $4096^2$ points, with $\bm{U}_\infty=(0.1,0)$ and $R=0.05$. A snapshot of the resulting vorticity field, and the corresponding pressure field, is shown in Fig.~\ref{fig:cylVortPress}.
To verify the accuracy of the spatial distribution of pressure, the surface-pressure coefficient ($C_P = 2P/\rho \|\bm{U}_\infty\|^2$) is plotted as a function of the azimuthal angle in Fig.~\ref{fig:CpVsTheta} ($\theta=0$ at the rear stagnation point).
The pressure distribution compares well with reference data from~\cite{Li2004}, at the three different time instances shown. To ensure that the velocity derivatives used for estimating the viscous forces (section~\ref{sec:viscForces}) are computed accurately, we examine the spatial distribution of vorticity on the lifted-surface of the cylinder (Fig.~\ref{fig:vortCyl}).
The distribution compares well with reference data from~\cite{Koumoutsakos1995} at two different time instances, for both $Re=550$ (Fig.~\ref{fig:vortRe550}) and $Re=1000$ (Fig.~\ref{fig:vortRe1000}). The temporal evolution of the pressure-induced and viscous contributions to the net drag force (Eqs.~\ref{eq:CdP} and~\ref{eq:CdNu}) is shown in Fig.~\ref{fig:CdCylinder}, and agrees reasonably well with reference data from~\cite{Koumoutsakos1995}.
The results discussed in this section suggest that the numerical algorithms described for determining pressure-induced and viscous forces work well for the case of rigid cylindrical objects.

%%%%%%%%%%%%%%%%%%%%%%%%%

\subsection{Impulsively started flow over a rigid streamlined object}
\begin{figure}
\centering
\includegraphics{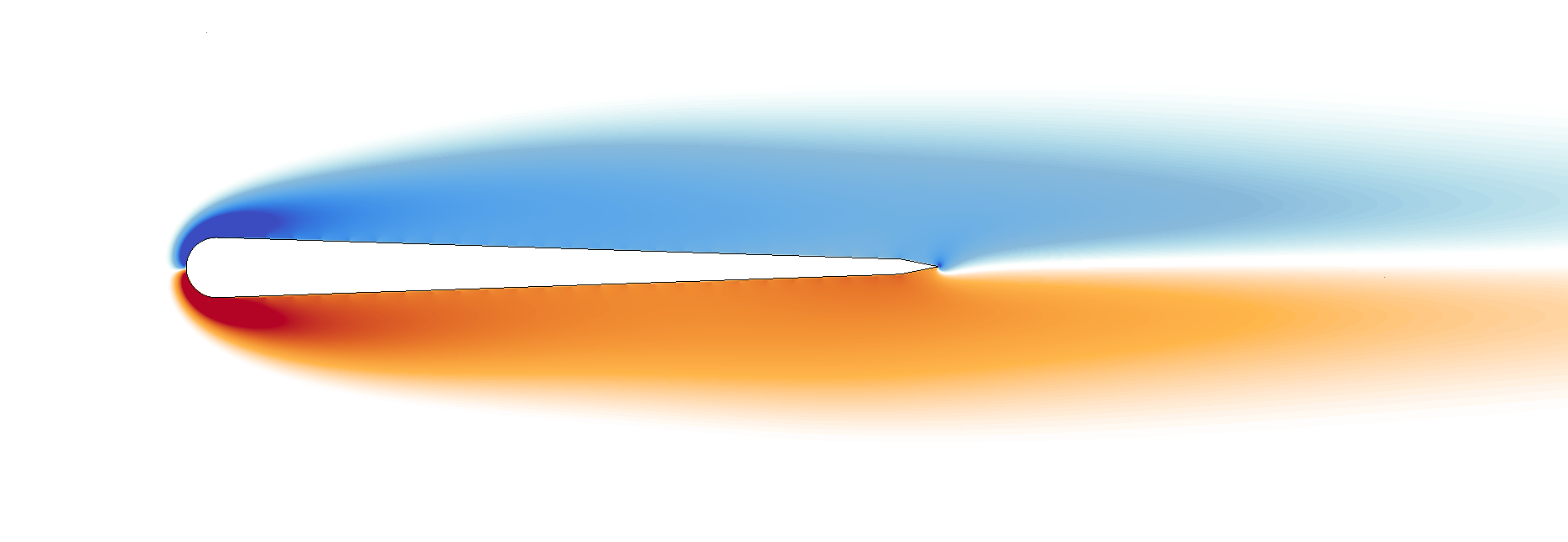}
\caption{Vorticity field around a rigid fish-shaped object in a uniform flow ($T=2$, $Re=400$). Regions of positive vorticity are colored in red, and regions with negative values are colored in blue.}
\label{fig:deadFishVort}
\end{figure}

To ensure that the surface-force computations perform well for non-cylindrical shapes, we examine impulsively started flow over a rigid streamlined object. The profile shape, shown in Fig.~\ref{fig:deadFishVort},
is based on a simplified 2D model of zebrafish~\cite{Carling1998,Kern2006,Gazzola2011}. The half-width of the profile is described as follows:
\begin{equation}
    w(s)= 
\begin{dcases}
    \sqrt{2w_hs-s^2} 					& 0\le s<s_b\\
    w_h-(w_h-w_t)\left(\dfrac{s-s_b}{s_t-s_b}\right) 	& s_b \le s < s_t \\
    w_t\dfrac{L-s}{L-s_t}				& s_t \le s \le L
\end{dcases}
\label{eq:fishWidth}
\end{equation}
where $s$ is the arc-length along the midline of the geometry, L is the body length, $w_h=s_b=0.04L$, $s_t=0.95L$, and $w_t=0.01L$. The relevant drag coefficient, the normalized time, and the Reynolds number are defined below:
\begin{subequations}
\begin{eqnarray}
Cd &=& \dfrac{(Drag)_P + (Drag)_\nu}{\rho \|\bm{U}_\infty\|^2 L/2}\\
T &=& tL/\|\bm{U}_\infty\| \\
Re &=& \dfrac{\|\bm{U}_\infty\| L}{\nu}
\end{eqnarray}
\end{subequations}

\begin{figure}
\centering
\includegraphics{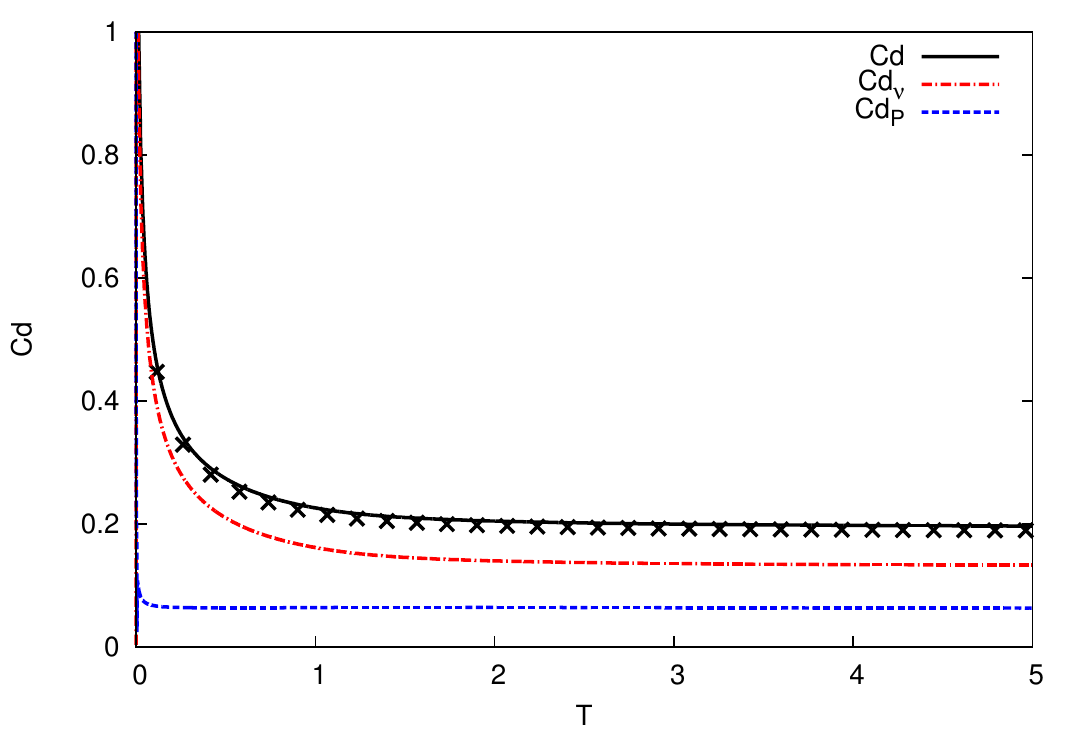}
\caption{Temporal evolution of pressure-drag coefficient ($Cd_p$) and viscous-drag coefficient ($Cd_\nu$) for flow over the fish-shaped object shown in Fig.~\ref{fig:deadFishVort}. Lines: computed using surface-forces, symbols: computed using the penalization algorithm (Eq.~\ref{eq:dragPenal}).}
\label{fig:CdDeadFish}
\end{figure}

Figure~\ref{fig:CdDeadFish}
shows the temporal evolution of drag coefficients for the body at a Reynolds number of 400, with $L=0.1$ and $\bm{U}_\infty=(0.1,0)$. The simulation was run on an adaptive grid with an effective resolution of $4096^2$ grid points. The viscous-drag coefficient ($Cd_\nu$) was obtained by evaluating $\bm{D}(\bm{u})$ on a lifted-surface, as described in Section~\ref{sec:viscForces}. The net drag coefficient (the solid line in Fig.~\ref{fig:CdDeadFish}) was computed as the sum of $Cd_\nu$ and $Cd_P$. The symbols in Fig.~\ref{fig:CdDeadFish} indicate drag determined by integrating the penalization term in Eq.~\ref{eq:penalNS} (with $\bm{u}_s = 0$)~\cite{Coquerelle2008}:
\begin{equation}
\bm{F}_{penal} = \iiint \lambda \chi  \bm{u} \ dV \quad \text{and} \quad Cd_{penal} = \dfrac{2\bm{F}_{penal}\cdot\bm{U}_\infty}{\rho \|\bm{U}_\infty\|^2L}
\label{eq:dragPenal}
\end{equation}

At the start of the simulation ($T\ll 1$), the static body experiences a large amount of drag, owing to the impulsively started flow. As the flow approaches steady state ($T>2$), the drag coefficients asymptote to constant values. We observe that the pressure-induced drag is small, even in the early stages of the simulation, which is a consequence of the streamlined shape of the body. This is not the case for blunt-shaped profiles (e.g., cylinders: Fig.~\ref{fig:CdCylinder}), where the pressure-induced drag (or form-drag) dominates throughout the simulation. Overall, the time-evolution of $Cd$ determined using surface-forces compares well with that computed using Eq.~\ref{eq:dragPenal}, which suggests that the numerical procedures outlined in Section~\ref{sec:numMeth} are suitable for use with rigid streamlined geometries.

%%%%%%%%%%%%%%%%%%%%%%%%%
\subsection{Self-propelled swimmers}

To validate the surface-force computations for dynamically deforming objects, simulations of self-propelled swimmers were performed by imposing time-varying deformations on 2D models of zebrafish~\cite{Carling1998,Kern2006,Gazzola2011}. The half-width of the swimmer's profile is defined by Eq.~\ref{eq:fishWidth}. The traveling-wave that describes the lateral displacement of the swimmer's midline is given as~\cite{Carling1998,Gazzola2011}:
\begin{equation}
y_s\left(s,t\right) = \dfrac{4}{33} \left(s+\dfrac{L}{32}\right)sin\left(2\pi\left( \dfrac{s}{L} - \dfrac{t}{T_p} \right)\right),
\end{equation}
where $T_p$ represents the tail-beat period imposed on the swimmer. Further details regarding the deformation and discretization of the profile may be found in~\cite{Gazzola2011}.

\begin{figure}
\centering
\includegraphics{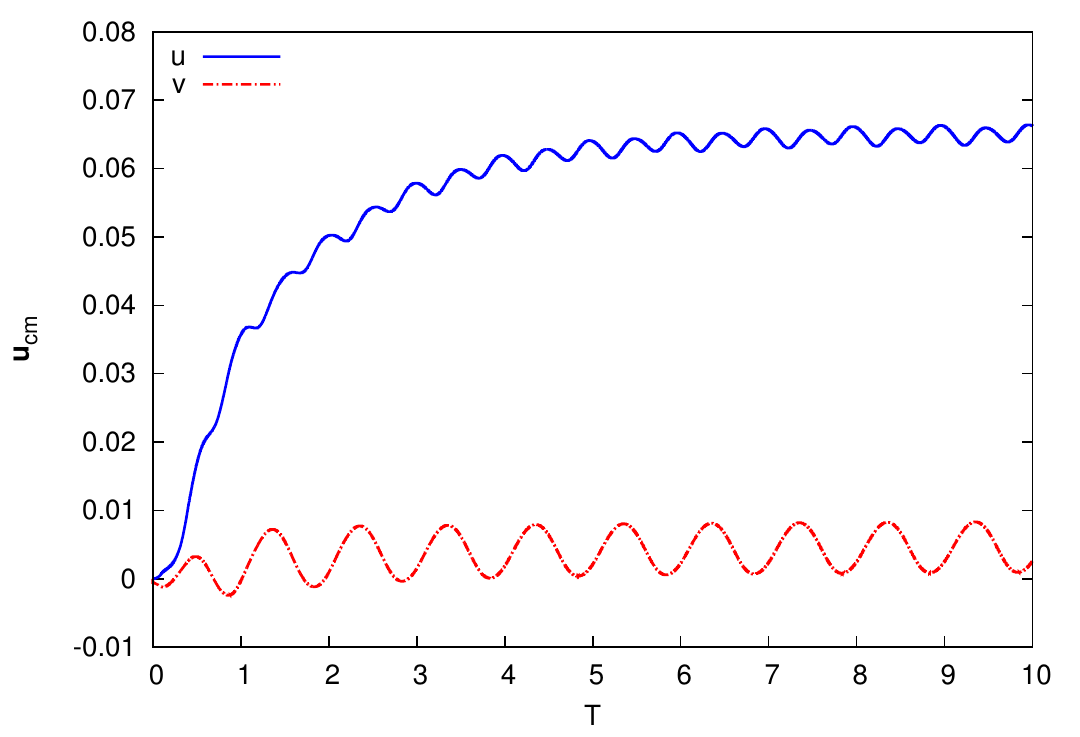}
\caption{The horizontal and vertical components of the center-of-mass-velocity for the swimmer.}
\label{fig:fishVel}
\end{figure}

\begin{figure}
        \centering
        \begin{subfigure}[b]{\textwidth}
                \centering
                \includegraphics{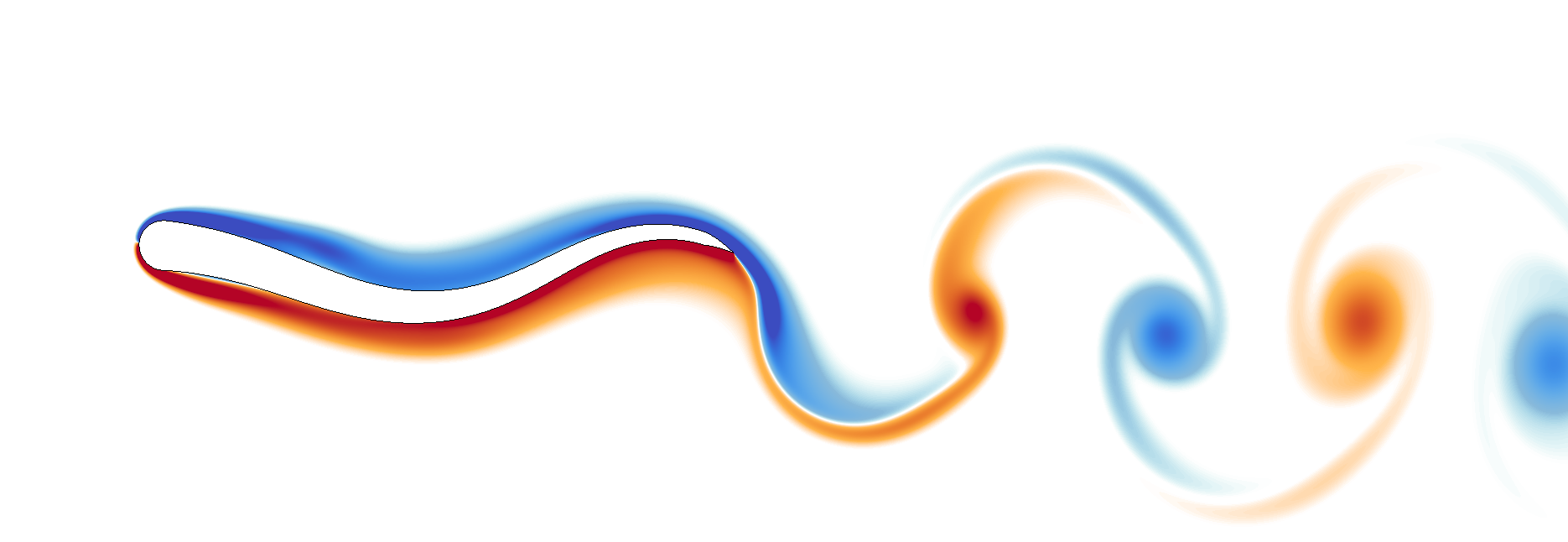}
                \subcaption{}
                \label{fig:fishFullVort}
        \end{subfigure}
        \begin{subfigure}[b]{\textwidth}
                \centering
                \includegraphics{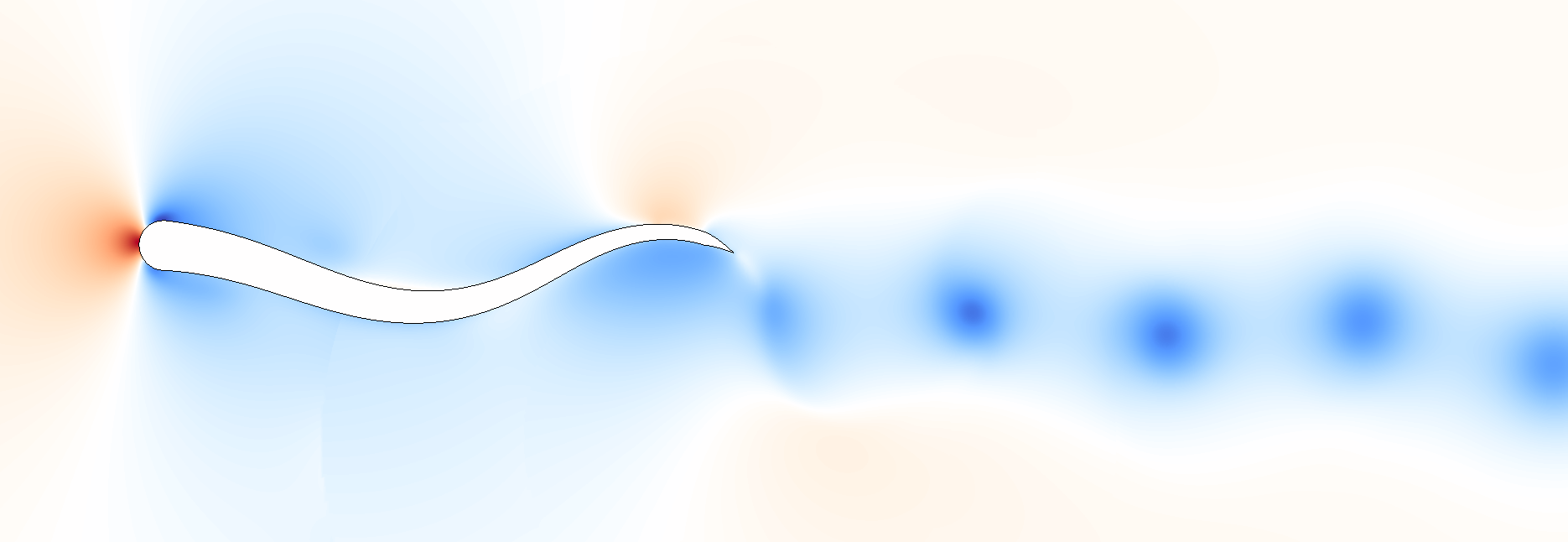}
                \subcaption{}
                \label{fig:fishFullPress}
        \end{subfigure}
        \caption{(\subref{fig:fishFullVort}) The vorticity field and (\subref{fig:fishFullPress}) the pressure field around the self-propelled swimmer. Regions of positive vorticity (and pressure) are colored in red, and regions with negative values are colored in blue.}
\label{fig:fishFull}
\end{figure}

The simulation of the swimmer was conducted at a Reynolds number of 4000, on an adaptive grid with an effective resolution of $8192^2$ grid points. This $Re$ value corresponds to the swimming of adult zebrafish, and is based on the length of the fish $L=0.1$, and the tail-beat period $T_p=1$ ($Re=L^2/T_p\nu$). The resulting time-evolution of the velocity of the center of mass of the swimmer ($\bm{u}_{cm}$) is shown in Fig.~\ref{fig:fishVel}. 
The swimmer accelerates from rest in the first few tail-beat periods ($T=t/T_p\le4$), before settling down to steady, periodically varying motion.
% The mean vertical velocity at the end of the simulation is non-zero as a consequence of the asymmetry introduced by the first tail-beat, which changes the inclination of the swimmer with respect to the horizontal axis.
The vorticity and pressure fields resulting from the simulation are shown in Fig.~\ref{fig:fishFull}.
The pressure field was computed by evaluating Eq.~\ref{eq:expandedFinal} with all grid nodes in the domain designated as target locations. A high pressure region develops in front of the head owing to the presence of a stagnation point. Acceleration of fluid around the head creates regions of low pressure on either side of the head, and the vortex-cores visible in Fig.~\ref{fig:fishFullVort} give rise to regions of low pressure in the fish's wake (Fig.~\ref{fig:fishFullPress}).

\begin{figure}
        \centering
        \begin{subfigure}[b]{\textwidth}
                \centering
                \includegraphics{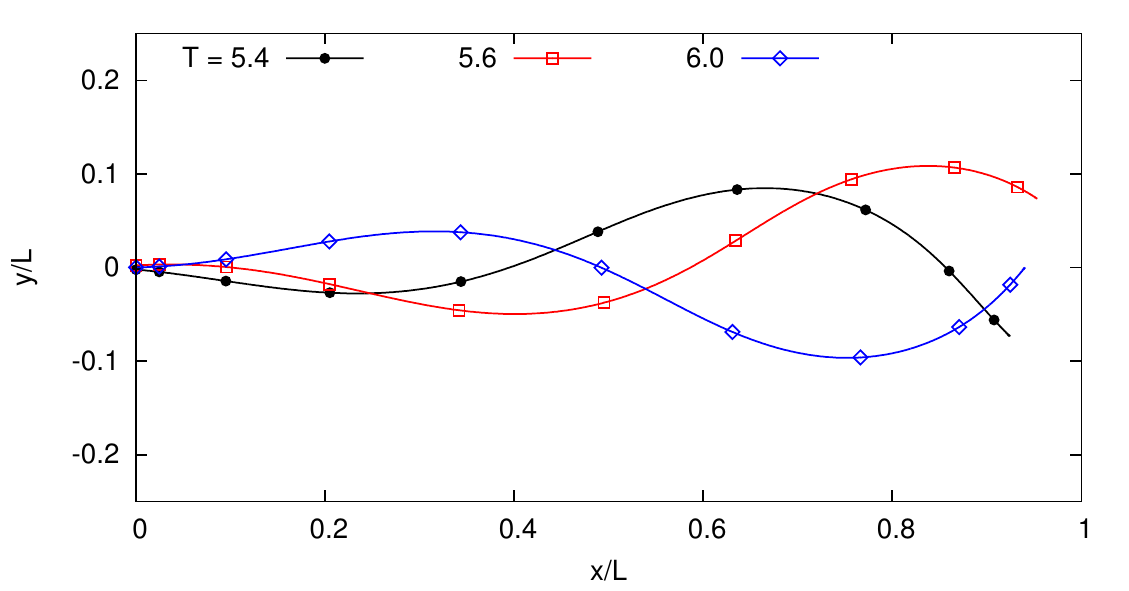}
                \subcaption{}
                \label{fig:fishKinematics}
        \end{subfigure}
        \begin{subfigure}[b]{\textwidth}
                \centering
                \includegraphics{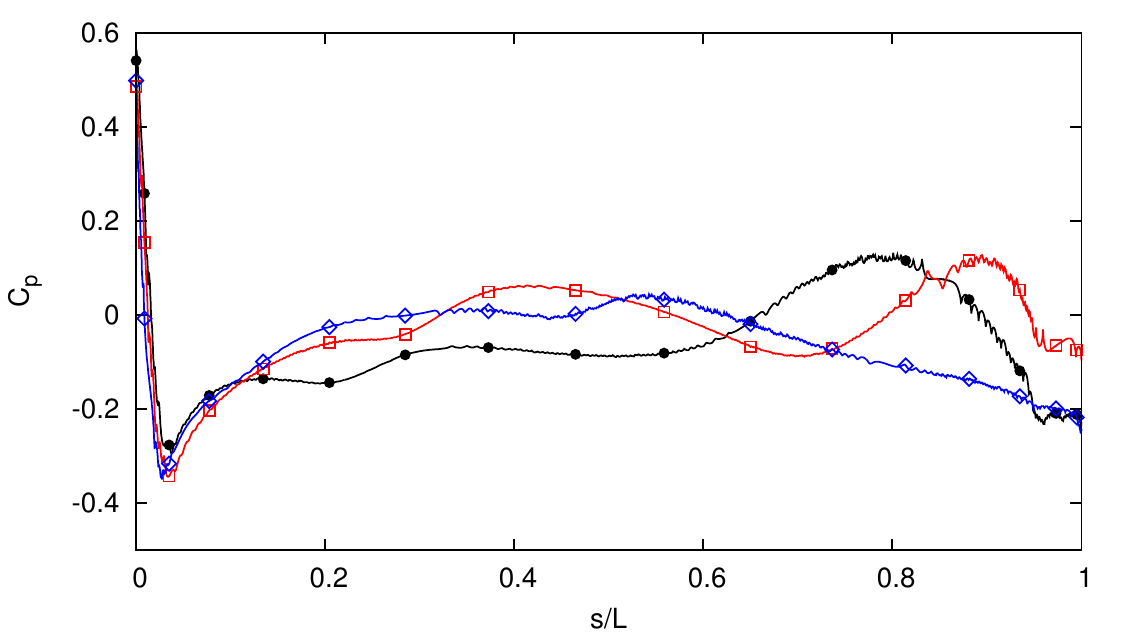}
                \subcaption{}
                \label{fig:fishSurfPress}
        \end{subfigure}
        \begin{subfigure}[b]{\textwidth}
                \centering
                \includegraphics{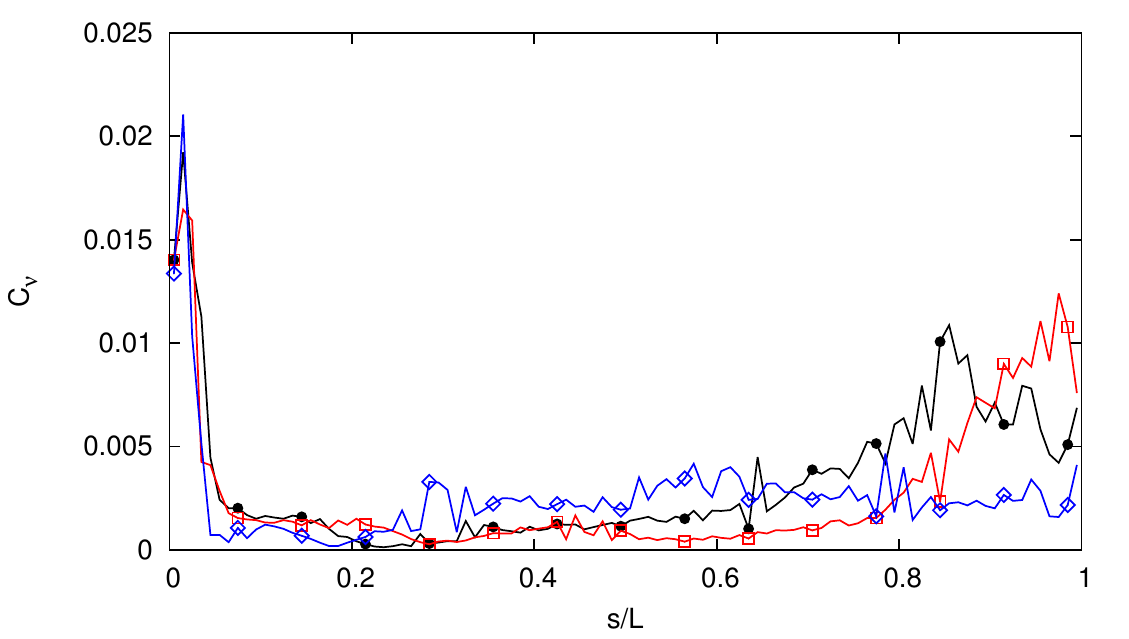}
                \subcaption{}
                \label{fig:fishSurfShear}
        \end{subfigure}
		\caption{(\subref{fig:fishKinematics}) Fish midline shape at various instances during a tail-beat period. Distribution of (\subref{fig:fishSurfPress}) pressure and (\subref{fig:fishSurfShear}) shear-traction magnitude on the right lateral surface, as a function of the midline arclength $s$. The curves in (\subref{fig:fishSurfShear}) have been averaged over intervals of length $0.01L$.}
		\label{fig:fishSurfPlots}
\end{figure}

In Fig.~\ref{fig:fishSurfPlots},
we examine the spatial distribution of pressure and shear-based quantities on the surface of the swimmer's body. Figure~\ref{fig:fishKinematics} depicts the shape of the fish-midline at three different instances during a tail-beat period. Figure~\ref{fig:fishSurfPress} shows the corresponding distribution of pressure coefficient on the right lateral surface of the fish's body. Figure~\ref{fig:fishSurfShear} depicts the magnitude of the shear (viscous) traction vector ($\|\bm{t}_\nu\|=\|2\mu\bm{D}\cdot\bm{n}\|$) on the right lateral surface. Both the pressure and the traction vector magnitude have been non-dimensionalized with $M/LT_P^2$ ($M$ represents the mass of the swimmer) to obtain the respective coefficients $C_p$ and $C_\nu$ shown in the plots. From Figs.~\ref{fig:fishSurfPress} and~\ref{fig:fishSurfShear}, we can surmise that the pointwise contribution from pressure is much larger ($\max(C_p)\approx0.5$) than the contribution from shear stress ($\max(C_\nu)\approx0.025$), which is expected for moderately large Reynolds numbers ($Re=4000$). The shear stress contribution is largest close to the tip of the head, followed by a pronounced drop along the midsection of the body, and a recovery towards the tail-end. The most dominant pressure contribution occurs at the head tip, followed by a persistent negative-$C_p$ region at the side of the head, and periodically varying $C_p$ along the rest of the body. The pointwise forces that emerge as a consequence of these stress distributions are depicted in an animation provided as part of the supplementary materials (Movie 1).

\begin{figure}
\centering
\includegraphics{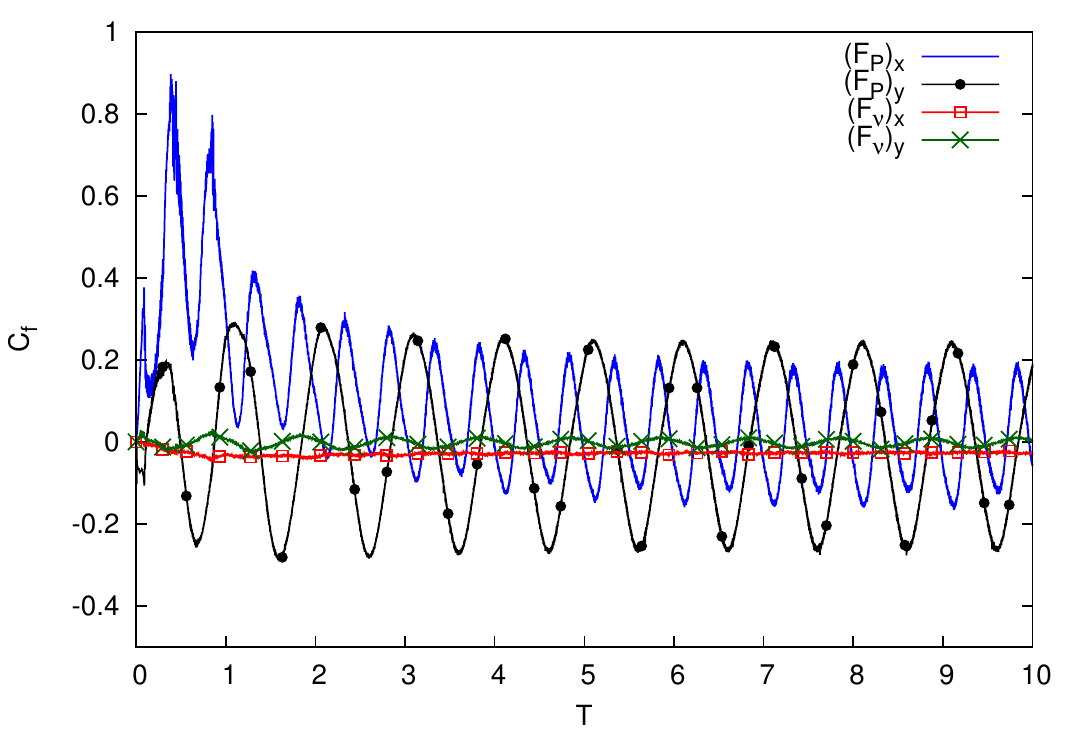} 
\caption{\label{fig:forcePlots}Time-evolution of the resultant surface-forces experienced by a self-propelled swimmer at $Re=4000$. The forces have been normalized by $ML/T_P^2$ to obtain the force coefficient $C_f$.}
\end{figure}

\begin{figure}
        \centering
        \begin{subfigure}[b]{\textwidth}
                \centering
                \includegraphics{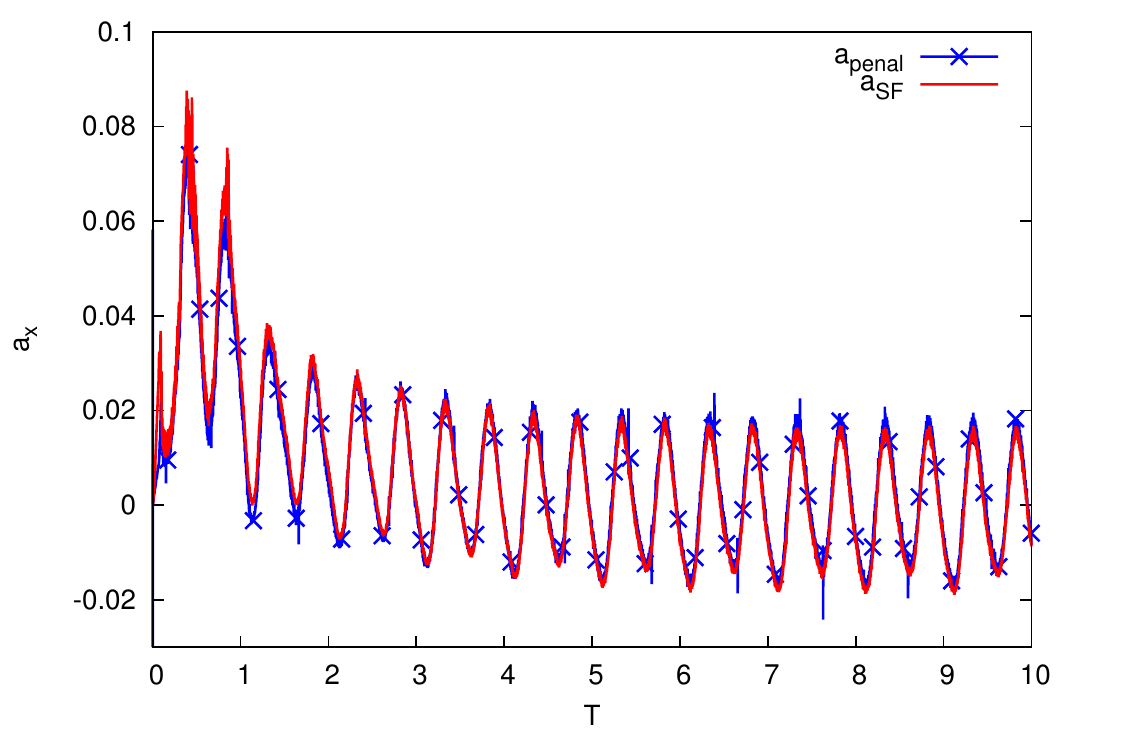}
                \subcaption{}
                \label{fig:aXfish}
        \end{subfigure}
        \begin{subfigure}[b]{\textwidth}
                \centering
                \includegraphics{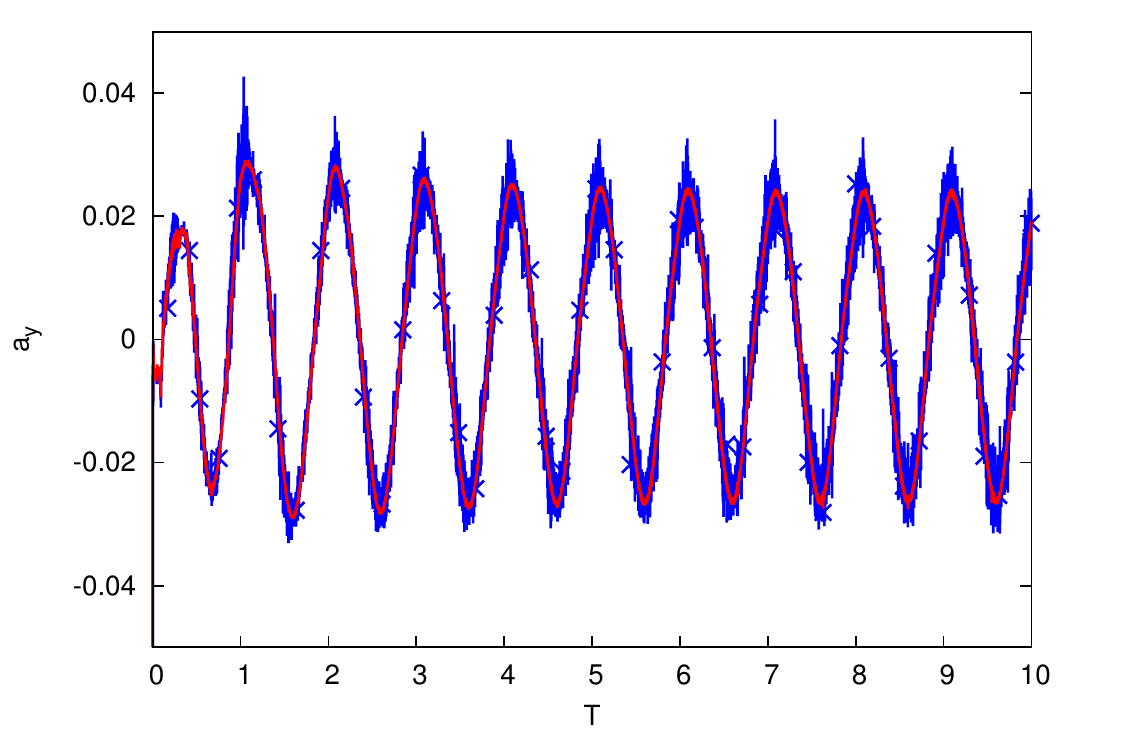}
                \subcaption{}
                \label{fig:aYfish}
        \end{subfigure}
		\caption{Comparison of (\subref{fig:aXfish}) horizontal acceleration and (\subref{fig:aYfish}) vertical acceleration of the self-propelled swimmer, computed using Eqs.~\ref{eq:aPenal} and~\ref{eq:aSF}.}
		\label{fig:accelPlots} 
\end{figure}

\begin{figure}
        \centering
        \begin{subfigure}[b]{\textwidth}
                \centering
                \includegraphics{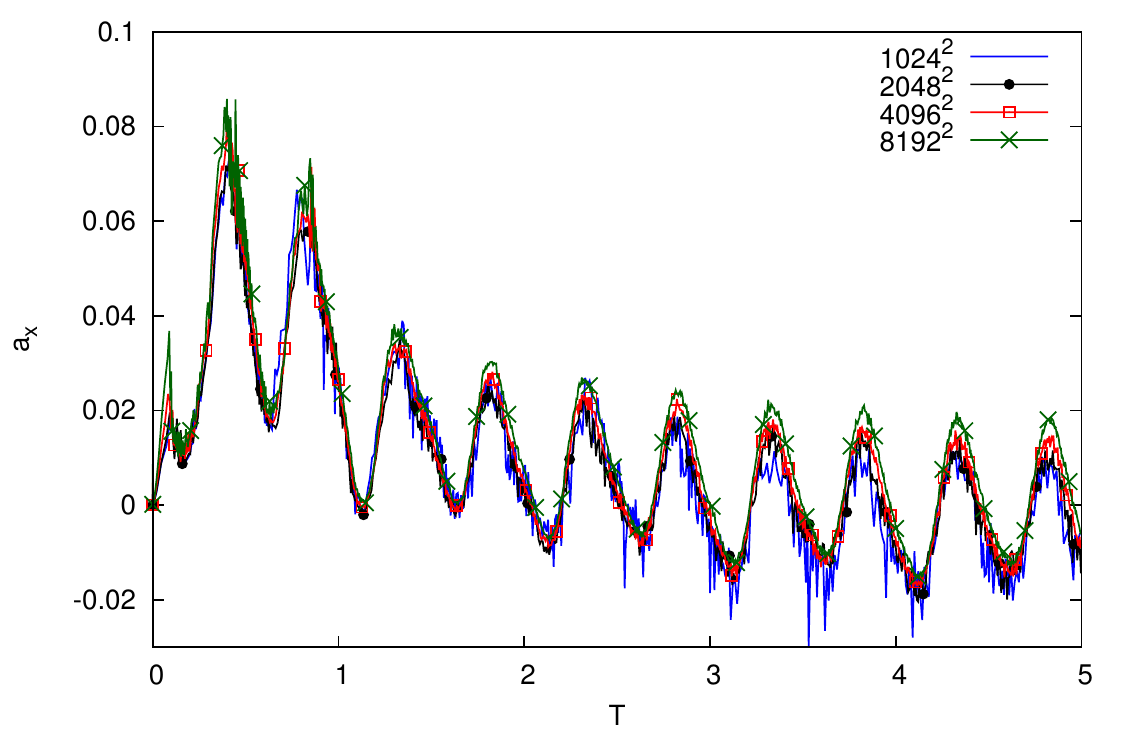}
                \subcaption{}
                \label{fig:aXres}
        \end{subfigure}
        \begin{subfigure}[b]{\textwidth}
                \centering
                \includegraphics{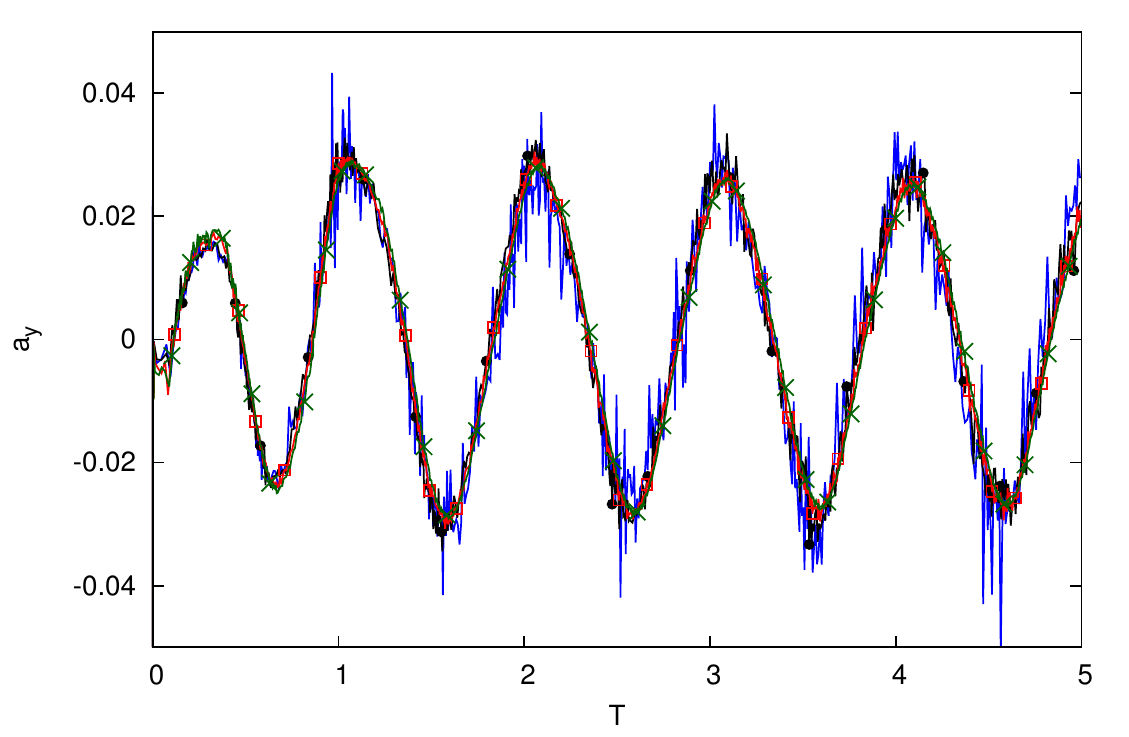}
                \subcaption{}
                \label{fig:aYres}
        \end{subfigure}
		\caption{(\subref{fig:aXres}) Horizontal acceleration and (\subref{fig:aYres}) vertical acceleration of the self-propelled swimmer, computed using Eq.~\ref{eq:aSF} for simulations conducted at four different grid resolutions.}
		\label{fig:accelPlotsGridConvg}
\end{figure}

The time-evolution of the resultant of the surface-forces (Eqs.~\ref{eq:FpressNet} and~\ref{eq:FviscNet}) is shown in Fig.~\ref{fig:forcePlots}.
We observe that the contribution of viscous forces is small compared to the contribution of pressure-induced forces, which matches our observation in Fig.~\ref{fig:fishSurfPlots}. The spikes in $(F_P)_x$ for $T<1$ suggest that the pressure distribution created by undulations of the swimmer plays a major role in accelerating the body from rest. For quantitative validation of the pressure-induced and viscous force computations, the horizontal ($x$) and vertical ($y$) components of net acceleration determined using surface-forces are compared to the acceleration predicted by the penalization algorithm. The acceleration at timestep `$n$' from the penalization algorithm is recovered from the object's center-of-mass-velocity using $2^{nd}$ order centered difference:
\begin{equation}
\bm{a}_{penal}^n = \dfrac{\bm{u}^{n+1}_{CM} - \bm{u}^{n-1}_{CM}} {2\Delta t}
\label{eq:aPenal}
\end{equation}
The acceleration corresponding to surface-forces is computed as follows:
\begin{equation}
\bm{a}_{SF}^n = \dfrac{\bm{F}_P^n + \bm{F}_\nu^n}{M},
\label{eq:aSF}
\end{equation}
where $M$ is the total mass of the solid object. The temporal evolution of $\bm{a}_{penal}$ and $\bm{a}_{SF}$ is shown in Fig.~\ref{fig:accelPlots}. 
The plots indicate that the net acceleration computed using surface-forces compares well with that determined using the penalization algorithm. The noise observed in the case of $\bm{a}_{penal}$ can be attributed to the use of finite-differences for computing the temporal derivative (Eq.~\ref{eq:aPenal}). The spikes in $a_x$ for $T<1$ correlate with the large $(F_P)_x$ values observed in Fig.~\ref{fig:forcePlots}. At later stages in the simulation, both $a_x$ and $a_y$ oscillate about a mean value of 0, which corresponds to the horizontal and vertical velocities approaching a steady mean value. Overall, the trends in $a_x$ and $a_y$ follow those observed for $(F_P)_x$ and $(F_P)_x$ in Fig.~\ref{fig:forcePlots}. To ensure that the algorithm works well on coarse grids, we compare the swimmer's net acceleration computed using surface-forces on four different grid resolutions (Fig.~\ref{fig:accelPlotsGridConvg}).
The acceleration exhibits increased levels of noise at lower resolutions, however, the values compare well to the data computed at higher resolutions. This result, as well as the good agreement between $\bm{a}_{penal}$ and $\bm{a}_{SF}$ in Fig.~\ref{fig:accelPlots}, confirms that the surface-forces computations described in Section~\ref{sec:numMeth} work well for the case of dynamically deforming solid objects. 

\section{Summary}
\label{sec:summary}
In this paper, we describe numerical procedures for determining flow-induced surface-forces on rigid and deforming bodies, when using vortex methods combined with Brinkman penalization. The pressure-Poisson equation is solved using a fast tree-code algorithm based on multipole expansions. Viscous forces are computed by evaluating the strain-rate tensor either on the object's surface, or on a lifted-surface, depending on whether a deforming or a rigid object is being simulated. Numerical tests involving impulsively started cylinders, a streamlined rigid fish-shaped body, and a self-propelled swimmer, are used to assess the efficacy of the method described. For the case of impulsively started cylinders, the spatial distribution of surface-pressure compares well with the results of benchmark simulations. Moreover, the temporal evolution of the pressure-induced and viscous drag coefficients shows good agreement with reference data. For streamlined fish-shaped profiles, drag measurements computed using surface-forces match those determined using the penalization algorithm. In simulations of self-propelled swimmers, the net unsteady acceleration calculated using the surface-force computations agrees well with the acceleration determined from the penalization algorithm. The tests presented indicate that the numerical procedures described are quite effective for determining pointwise surface-forces on complex, temporally evolving geometries. Future work involves the extension of this method to three-dimensional flows.

\acks
We gratefully acknowldge support by the European Research Council Advanced Investigator Award, and the Swiss National Science Foundation Sinergia Award (CRSII3\_147675). This work utilized computational resources granted by the Swiss National Supercomputing Centre (CSCS) under project ID `s436'.

\appendix
\section{Numerical solution of the pressure Poisson equation}
\label{sec:PoissonNumerics}
\subsection{The Green's function}
There are several ways of numerically solving the Poisson's equation shown in Eq.~\ref{eq:PoissonFinal}, for instance, using Fourier-based fast Poisson solvers~\cite{Hockney1965}, using multigrid methods~\cite{Stuben1982, Gupta1997}, or using tree-code algorithms~\cite{Appel1985, Rokhlin1985,Barnes1986,Greengard1987}. For solving the Poisson's equation on the multi-resolution adaptive grids~\cite{Rossinelli2015} used in our simulations, we use the tree-code algorithm, which requires knowledge of the Green's function (or the fundamental solution) for the Laplacian.

The free-space Green's function $G\left(\bm{x}, \bm{x}'\right)$ for the Laplacian is the solution of the following equation, with the Neumann boundary condition specified at infinity:
\begin{eqnarray}
\nabla^2 G\left(\bm{x}, \bm{x}'\right)  &=& \delta\left(\bm{x} - \bm{x}'\right)\label{eq:greenDelta}\\
\nabla G\left(\bm{x}, \bm{x}'\right)  &=& 0 \quad \text{on } \quad dS
\label{eq:greenNeumann}
\end{eqnarray}
Physically, the Neumann boundary condition in Eq.~\ref{eq:greenNeumann} corresponds to the pressure-forces being zero at infinity. The Green's function that satisfies Eqs.~\ref{eq:greenDelta} and~\ref{eq:greenNeumann} in two-dimensions is:
\begin{equation}
G\left(\bm{x}, \bm{x}'\right) = \dfrac{1}{2\pi}\log\left(\|\bm{x} - \bm{x}'\|\right)
\label{eq:green2D}
\end{equation}
This fundamental solution can be used to solve the inhomogeneous pressure-Poisson equation, by computing its convolution with the forcing term on the right hand side of Eq.~\ref{eq:PoissonFinal}:
\begin{equation}
P\left(\bm{x}\right) = \int G\left(\bm{x}, \bm{x}'\right) \left(-\rho\left(\nabla\bm{u}^T:\nabla\bm{u}\right) + \rho\lambda'\nabla\cdot\left(\chi\left(\bm{u}_s-\bm{u}\right)\right)\right) d\bm{x}'
\label{eq:greenConv}
\end{equation}
where $\bm{u}$, $\bm{u}_s$, and $\chi$ are functions of $\bm{x}'$. Discretizing the integral in Eq.~\ref{eq:greenConv} yields the following summation:
\begin{equation}
P\left(\bm{x}\right) = \sum_{\bm{x}'} G\left(\bm{x}, \bm{x}'\right)f(\bm{x}') h^2(\bm{x}')
\label{eq:greenSum}
\end{equation}
where $h^2$ represents the area of the grid-cell at location $\bm{x}'$ , and
\begin{equation}
f(\bm{x}') = -\rho\left(\nabla\bm{u}(\bm{x}')^T:\nabla\bm{u}(\bm{x}')\right) + \rho\lambda'\nabla\cdot\left(\chi(\bm{x}')\left(\bm{u}_s-\bm{u}(\bm{x}')\right)\right)
\end{equation}
Note that $h$ is a function of $\bm{x}'$, as the cell size may vary spatially when using adaptive grids.

Equation~\ref{eq:greenSum} sums up contributions from source-points located throughout the domain (hence the summation over $\bm{x}'$), to compute the pressure at location $\bm{x}$. Evaluating this sum directly is computationally intractable even for problems that are relatively modest in size, and a tree-code~\cite{Barnes1986} combined with multipole expansions~\cite{Greengard1987} is used to speed-up the computation.

%%%%%%%%%%%%%%%%%%%%%%%%%%%%%%%%%%%
\subsection{Multipole Expansion}
Tree-codes and multipole methods have been used extensively for solving `N-body' problems involving gravitational and electric fields~\cite{Barnes1986, Greengard1987, Greengard1990}, and for fluid-simulations based on vortex dynamics~\cite{Pepin1990, Koumoutsakos1995,Gazzola2011}. These methods can be used to reduce the computational complexity of the summation in Eq.~\ref{eq:greenSum} from $O(N^2)$ to $O(N\log N)$~\cite{Barnes1986}, or to $O(N)$~\cite{Greengard1987,Greengard1990}, depending on the specific algorithm used. The computational savings result from the combined use of a hierarchical quadtree data structure, and a truncated series expansion for $G\left(\bm{x}, \bm{x}'\right)$. To obtain the multipole expansion for the logarithmic function in Eq.~\ref{eq:green2D}, the coordinates $\bm{x}$ and $\bm{x}'$ are expressed as complex numbers $\bm{z}  = x + iy$ and $\bm{z}' = x' + iy'$. To facilitate series expansion of the resulting expression ($\log\left(\|\bm{z}-\bm{z}'\|\right)$), we express the difference $\bm{z}-\bm{z}'$ in terms of polar coordinates $r e^{i\theta}$ (where $r = \|\bm{z}-\bm{z}'\|$ and $\theta = \arg\left(\bm{z}-\bm{z}'\right)$). The complex logarithm $\log\left(\bm{z}-\bm{z}'\right)$ and the real-valued function in Eq.~\ref{eq:green2D} are related as follows:
\begin{eqnarray}
\log\left(\bm{z}-\bm{z}'\right) &=& \log\left(r\right) + i\theta\\
\Rightarrow \log\left(\|\bm{x} - \bm{x}'\|\right)&=& \Re\{\log\left(\bm{z}-\bm{z}'\right)\}
\label{eq:logRealToComplex}
\end{eqnarray}

Equation~\ref{eq:logRealToComplex} suggests that $G\left(\bm{x}, \bm{x}'\right)$ can be represented as the real part of the series expansion for $\log\left(\bm{z}-\bm{z}'\right)$:
\begin{eqnarray}
\log\left(\bm{z}-\bm{z}'\right) &=& \log\left(\bm{z}\left(1-\dfrac{\bm{z}'}{\bm{z}}\right)\right) \\
&=& \log\left(\bm{z}\right) - \sum_{k=1}^{\infty}\dfrac{1}{k}\left(\dfrac{\bm{z}'}{\bm{z}}\right)^k \quad \text{iff} \quad  \|\bm{z}'\|<\|\bm{z}\|
\label{eq:logExpansion}
\end{eqnarray}
Assuming that the computation for $P\left(\bm{z}\right)$ involves $N$ source particles, we combine Eqs.~\ref{eq:green2D}, \ref{eq:greenSum}, \ref{eq:logRealToComplex}, and~\ref{eq:logExpansion} to get:
\begin{equation}
P\left(\bm{z}\right) = \dfrac{1}{2\pi}\sum_{i=1}^{N} f(\bm{z}_i') h^2 (\bm{z}_i') \Re\left\{ \log\left(\bm{z}\right)\right\} - \dfrac{1}{2\pi}\sum_{i=1}^{N} f(\bm{z}_i') h^2(\bm{z}_i') \Re\left\{\sum_{k=1}^{\infty}\dfrac{1}{k}\left(\dfrac{\bm{z}_i'}{\bm{z}}\right)^k \right\}
\label{eq:expanded1}
\end{equation}
The summation in the first term in Eq.~\ref{eq:expanded1} simplifies to a constant value, which represents the total contribution $S$ from all source-points. The summation signs in the second term are interchanged, and the summation over $k$ is truncated to $p$ terms:
\begin{subequations}
\begin{eqnarray}
P\left(\bm{z}\right) &=& \dfrac{S}{2\pi} \Re\left\{ \log\left(\bm{z}\right)\right\} - \sum_{k=1}^{p}\Re\left\{\dfrac{\alpha_k}{\bm{z}^k} \right\}
\label{eq:pressTruncated}\\
S &=& \sum_{i=1}^{N} f(\bm{z}_i') h^2 (\bm{z}_i') \\
\alpha_k &=& \sum_{i=1}^{N} f(\bm{z}_i')  h^2(\bm{z}_i')\dfrac{ \bm{z}_i^{'k} }{k}
\label{eq:alphaK}
\end{eqnarray}
\label{eq:expandedFinal}
\end{subequations}
Note that $S$ and the coefficients $\alpha_k$ are evaluated and stored in a single pass through the quadtree, as they do not depend on $\bm{z}$. The choice of truncation parameter `$p$' depends on the desired level of accuracy~\cite{Greengard1987}.

\begin{figure}
\centering
\includegraphics{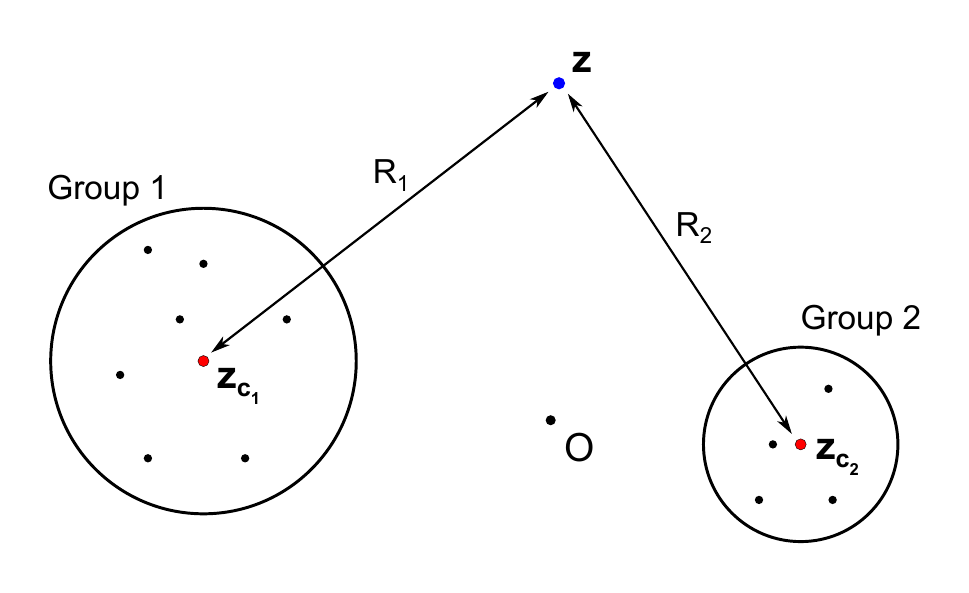}
\caption{Schematic showing pressure source-points (black dots) grouped together into two distinct clusters. $R_j$ denotes the distance from the center of the clusters $\left(\bm{z}_{c_j}\right)$ to the point of pressure evaluation $\left(\bm{z}\right)$. $O$ represents the origin in the complex plane.}
\label{fig:multipole}
\end{figure}

The multipole expansion in Eq.~\ref{eq:expandedFinal} and the hierarchical quadtree allow us to group source-points together, as depicted in Fig.~\ref{fig:multipole}.
Each group is perceived as a single source-point (the red dots located at $\bm{z}_{c_j}$ in Fig.~\ref{fig:multipole}) at sufficiently large distances~\cite{Appel1985,Barnes1986, Greengard1987,  Greengard1990}. Due to the linear nature of Eq.~\ref{eq:greenSum}, the contributions from individual groups can be combined to obtain $P\left(\bm{z}\right)$:
\begin{equation}
P\left(\bm{z}\right) = P\left(\bm{z}\right)_{Group_1} + P\left(\bm{z}\right)_{Group_2} + P\left(\bm{z}\right)_{Group_3} + \cdots
\end{equation}
The group contributions $P\left(\bm{z}\right)_{Group_j}$ can be evaluated by replacing $\bm{z}$ with $\left(\bm{z}-\bm{z}_{c_j}\right)$, and $\bm{z}'$ with $\left(\bm{z}'-\bm{z}_{c_j}\right)$ in Eq.~\ref{eq:expandedFinal}, and by limiting the summation over `$i$' to the sources contained within each group.

The multipole expansions for parent-clusters (i.e., large clusters composed of smaller clusters) can be generated by combining their childrens' expansions. This reduces the computational cost, since we no longer need to repeatedly evaluate the contributions from all source-points enclosed within larger clusters. The multipole expansion of a child-cluster centered at $\bm{z}_{c_j}$ can be shifted to the appropriate parent-cluster's center at $\bm{z}_{p}$, using an exact formula~\cite{Greengard1987}:
\begin{eqnarray}
P\left(\bm{z}\right)_{Group_j} &=& \dfrac{S_j}{2\pi} \Re\left\{ \log\left(\bm{z}-\bm{z}_{c_j}\right)\right\} - \sum_{k=1}^{p}\Re\left\{\dfrac{\alpha_k}{\left(\bm{z} - \bm{z}_{c_j}\right)^k} \right\}\\
&=&\dfrac{S_j}{2\pi} \Re\left\{ \log\left(\bm{z}-\bm{z}_{p}\right)\right\} - \sum_{k=1}^{p}\Re\left\{\dfrac{\beta_k}{\left(\bm{z}-\bm{z}_{p}\right)^k} \right\}
\label{eq:expandedShift}
\end{eqnarray}
where
\begin{equation}
\beta_k = -\dfrac{S_j \left(\bm{z}_{c_j} - \bm{z}_p\right)^k}{k} + \sum_{l=1}^{k} \alpha_l \left(\bm{z}_{c_j} - \bm{z}_p\right)^{k-l}{ k-1 \choose l-1}
\label{eq:expandedBeta}
\end{equation}
Starting from the finest level in the quadtree, Eqs.~\ref{eq:expandedShift} and~\ref{eq:expandedBeta} can be used to generate the expansions for all parent nodes recursively. Further details regarding the implementation of tree-codes may be found in~\cite{Barnes1986,Greengard1987,Greengard1990}.

\bibliography{surfForce_ArXiv}

\end{document}